# Chemical Reaction Rate Coefficients from Ring Polymer Molecular Dynamics: Theory and Practical Applications

Yury V. Suleimanov*
*Computation-based Science and Technology Research Center, Cyprus Institute, 20 Kavafi Street, Nicosia 2121, Cyprus*
*Department of Chemical Engineering, Massachusetts Institute of Technology, Cambridge, Massachusetts 02139, United States*
**Email: y.suleymanov@cyi.ac.cy; ysuleyma@mit.edu, +35722208721*

F. Javier Aoiz*
Departamento de Química Física I, Facultad de CC. Químicas, Universidad Complutense de Madrid, 28040 Madrid, Spain
**Email: aoiz@quim.ucm.es, +34913944126*

Hua Guo*
*Department of Chemistry and Chemical Biology, University of New Mexico, Albuquerque, New Mexico 87131, United States*
**Email: hguo@unm.edu, Phone number: +15052771716*

**ABSTRACT**
This Feature Article presents an overview of the current status of Ring Polymer Molecular Dynamics (RPMD) rate theory. We first analyze the RPMD approach and its connection to quantum transition-state theory. We then focus on its practical applications to prototypical chemical reactions in the gas phase, which demonstrate how accurate and reliable RPMD is for calculating thermal chemical reaction rate coefficients in multifarious cases. This review serves as an important checkpoint in RPMD rate theory development, which shows that RPMD is shifting from being just one of recent novel ideas to a well-established and validated alternative to conventional techniques for calculating thermal chemical rate coefficients. We also hope it will motivate further applications of RPMD to various chemical reactions.

*: corresponding authors

## 1. INTRODUCTION

A key objective in modern physical chemistry is to predict rates of chemical reactions, which depend on rate coefficients, usually as a function of temperature.[1] The challenge of theoretical kinetics is how to determine the rate coefficient accurately and efficiently from first principles.

Rate coefficients for bimolecular reactions are determined by intrinsic molecular interactions between the reactants. Such interactions can be often described within the Born-Oppenheimer (BO) approximation, which separates the electronic motion from the much slower nuclear motion.[2] In this Feature Article, we restrict our discussion to chemical reactions within the BO approximation, ignoring non-adiabatic effects. The BO approximation assumes that the interaction between two reactants is governed by the adiabatic potential energy surface (PES), which is the electronic energy (plus nuclear repulsion) as a function of nuclear coordinates. Nuclear dynamics on an adiabatic PES, which is characterized by the nuclear Schrödinger equation, yields the thermal rate coefficient as an average of the energy-dependent reactivity (reaction cross sections or cumulative reaction probabilities) over the Boltzmann distribution at a given temperature.

This quantum dynamical approach based on solving the nuclear Schrödinger equation requires (1) highly accurate *ab initio* calculations of the electronic energy at a large number of nuclear geometries, (2) a faithful representation of the global PES based on the *ab initio* points in the relevant nuclear configuration space, and (3) accurate quantum nuclear dynamics on the PES. Many advances have been made of late in all three areas.[3-7] Indeed, this approach has proven extremely accurate for reactive systems containing small numbers of atoms.[8-10] However, the numerical cost increases exponentially with the number of atoms, ruling out its application to larger systems. A less reliable approximation is to replace the quantum mechanical (QM) description of the nuclear dynamics with a Newtonian model. While the dynamical effects are explicitly taken into consideration, albeit classically, the quasi-classical trajectory (QCT) model ignores important quantum effects such as tunneling, zero-point energy conservation, and resonances. As a result, its reliability is sometimes uncertain, particularly at low energies and temperatures.[11,12]

It has long been realized that for a reaction between two reactants, the rate coefficient is largely determined by the overcoming of a transition state. The concept of the transition state was first proposed by Erying, Evans, Polanyi, and Wigner in 1930s as an activated complex, which once reached will dissociate irreversibly to products. While this idea is concise and appealing, the precise location of the transition state is neither easily definable theoretically nor measurable experimentally. Nonetheless, this concept laid the foundation for the enormously popular and successful transition-state theory (TST).[13-19] Within TST, the rate coefficient is proportional to the reaction flux passing through a dividing surface, through which the flux is irreversible. The flux can be computed by simply counting the density of states in the reactant asymptote and at the transition state, which is quite efficient as it requires neither the global PES nor QM calculations. In reality, however, the dividing surface is very difficult to define accurately. Dynamical factors near the transition state might induce recrossing of the dividing surface, thus reducing the reactivity. A solution of the problem is to optimize variationally the location of the dividing surface by minimizing the crossing, but this may not eliminate completely the recrossing effect. Additionally, quantum effects such as tunneling can be taken into account using approximate semi-classical approaches, such as the quantum instanton theory.[16] While the modern versions of TST have become a powerful tool to study chemical kinetics, particularly for systems involving heavy atoms and at high temperatures,[20] the accuracy of the computed rate coefficients depends on how exactly the quantum and dynamical effects are treated.

Despite the progress, there are still several issues in theoretical kinetics for which improvements are needed. One is the influence of anharmonicity, which is often ignored in conventional TST calculations.[21,22] Another is the multi-dimensional tunneling, which is difficult to be accurately treated locally within the TST assumption, particularly in the deep tunneling regime.[20,23] Finally, as commented on above, recrossing at the dividing surface, which is intrinsically a dynamical effect, requires a more rigorous treatment. These problems could result in significant errors in TST calculations under various conditions.[21-24]

The recently proposed Ring-Polymer Molecular Dynamics (RPMD) approach[25,26] offers a promising solution to all three problems outlined above. While introduced in an *ad hoc* fashion, RPMD takes advantage of the isomorphism between the statistical properties of a quantum system and those of a fictitious classical ring polymer consisting of harmonically connected beads, which allows the approximation of the real-time evolution of the quantum system by classical trajectories. As a result, this approximate quantum theory is highly efficient and scales favorably with dimensionality of the system. On the other hand, it preserves the quantum Boltzmann distribution and time-reversal symmetry.[26] It has emerged as an efficient and reasonably accurate method to include quantum effects in time-correlation functions.[25] The RPMD rate theory has several additional desirable features: (i) it becomes exact at the high temperature limit, where the ring polymer collapses to a single bead;[26] (ii) it preserves the zero-point energy (ZPE) along the reaction path; (iii) it has as a well-defined short-time limit[27-30] that offers an upper bound on the RPMD rate coefficient, in the same way as classical transition state theory provides an upper bound on the classical rate coefficient; (iv) it provides the exact solution for tunneling through a parabolic barrier;[31] and (v) it does not depend on the choice of the dividing surface and provides a parameter-free rate coefficient.[32] As discussed below in more detail, the static and dynamic factors counter-balance to make the location of the dividing surface inconsequential. This is a highly desirable feature because an optimal choice of the dividing surface in the multidimensional space is often very difficult.

Since its inauguration, the RPMD rate theory has been applied to many prototypical reactions[21,23,24,31-50] and the comparison with exact quantum dynamical results and various quantum dynamics approximations has been very promising.[21,23,31-34,37-42,50] A general-purpose software package (RPMDrate) has been written by one of the authors (Y.V.S.) and distributed publically.[51] We note in passing here that applications of RPMD theory in condensed phase problems are out of the scope of this review and will thus be discussed with only cursory attention.

In the remainder of this Feature article, we first discuss the formal aspects of the RPMD rate theory, placing an emphasis on its connection to various flavors of the quantum transition-state theory. It is shown that the RPMD formulation, although derived in a heuristic way, offers a rather reliable description of the reaction rate coefficient and is exact in some limiting cases. It also provides a quite accurate characterization of the tunneling event in reactions, even in the so-called "deep tunneling" region, because of the ability of RPMD to describe the correct quantum Boltzmann statistics. This feature is particularly important for kinetics at low temperatures. The formal discussion is followed by a detailed description of the specific implementation of the general RPMD theory for rate calculations. In addition, the selection of parameters necessary to converge the RPMD calculations is also discussed. Finally, the performance of the RPMD rate theory is assessed by comparing the results with the existing QM calculations for small bimolecular reactions and with experiment. Particular attention is given to activated reactions involving the transfer of light atoms, in which quantum effects such as zero-point energies, tunneling, and anharmonicity are amplified. Discussion on the applicability of the RPMD rate theory to complex-forming reactions is also presented. These investigations of prototypical reactions with various mass combinations and energetics allowed extensive validation of the method and demonstrated clearly that the RPMD approach is accurate, efficient, and superior to many alternative theoretical methods for computing rate coefficients. The review is cumulated with prospects of the RPMD rate theory in future applications.

## 2. RING-POLYMER MOLECULAR DYNAMICS RATE THEORY

In what follows, we give the key equations of the RPMD rate-theory, and explain why they give a good approximation to the exact quantum rate coefficient for a large class of chemical reactions. It is sufficient to consider a one-dimensional system, with the following classical Hamiltonian

$$H(p,q) = \frac{p^2}{2m} + V(q), \tag{1}$$

where $V(q)$ is some potential energy curve (not necessarily symmetric), which tends to become constant in the limits $q \to -\infty$ (reactants) and $q \to +\infty$ (products). Generalization of the theory to multiple dimensions is straightforward, as will be discussed in Section 3.

### 2.1. The RPMD rate coefficient

To start the discussion, we first define the ring-polymer Hamiltonian[26]

$$H_N(\mathbf{p},\mathbf{q}) = \sum_{i=1}^{N} \frac{p_i^2}{2m} + U_N(\mathbf{q}), \tag{2}$$

in which $\mathbf{q} = \{q_1, \ldots, q_N\}$ are a set of $N$ replicas of the system coordinate $q$, $\mathbf{p} = \{p_1, \ldots, p_N\}$ are the conjugate momenta, and

$$U_N(\mathbf{q}) = \sum_{i=1}^{N} \frac{m(q_{i+1} - q_i)^2}{2(\beta_N \hbar)^2} + V(q_i), \tag{3}$$

in which $\beta_N = \beta / N$, $\beta = 1/k_B T$ and the $\boldsymbol{q}$ satisfy the cyclic boundary condition $q_{i\pm N} = q_i$. Clearly $U_N(\mathbf{q})$ has the form of a classical ring-polymer potential, consisting of $N$ "beads" connected by harmonic springs.

The RPMD rate coefficient $k_{\mathrm{RP}}(T)$ is obtained by applying (standard) classical rate-theory within the (fictitious) extended phase-space (**p**,**q**), which gives[31,32]

$$k_{\mathrm{RP}}(T)Q(T) = \lim_{t\to\infty} C_{\mathrm{fs}}(t,T) \quad (4)$$

where $Q(T)$ is the reactant partition function, and $C_{\mathrm{fs}}(t,T)$ is the RPMD flux-side time-correlation function:

$$C_{\mathrm{fs}}(t,T) = \lim_{N\to\infty} \frac{1}{(2\pi\hbar)^N} \int d\mathbf{p} \int d\mathbf{q} e^{-\beta_N H_N} \delta\left[f(\mathbf{q})\right] \dot{f}(\mathbf{q}) h\left[f(\mathbf{q}_t)\right], \quad (5)$$

in which $\mathbf{q}_t$ denotes the positions of the ring-polymer coordinates at time $t$, $f(\mathbf{q}) = 0$ defines the dividing surface (*i.e.*, a function of **q** that separates the reactants from the products), and

$$\dot{f}(\mathbf{q}) = \sum_{i=1}^{N} \frac{\partial f(\mathbf{q})}{\partial q_i} \frac{p_i}{m} \quad (6)$$

is the $t \to 0_+$ flux through $f(\mathbf{q}) = 0$ (and $h(x)$ denotes the Heaviside step-function, and we introduce the notation $\int d\mathbf{q} = \int_{-\infty}^{\infty} dq_1 \ldots \int_{-\infty}^{\infty} dq_N$ used throughout). Because $k_{\mathrm{RP}}(T)$ is a classical rate (in an extended space), it satisfies the important property of being *independent of the position of the dividing surface* $f(\mathbf{q}) = 0$.[32]

The classical dynamics of the ring-polymers was initially introduced in an *ad hoc* manner.[26] In recent attempts to derive RPMD, its connection to the exact quantum Kubo-transformed time correlation function *via* a Boltzmann conserving "Matsubara dynamics" has been demonstrated using a heuristic approach (with explicit terms that are discarded).[52,53] "Matsubara dynamics" considers evolution of the low-frequency, smooth "Matsubara" modes of the path integral,[54] although it suffers from the sign problem and is not presently amenable to practical computation on realistic systems. However, if the momentum contour is moved into the complex plane in order to make the quantum Boltzmann distribution real, a complex Liouvillian arises, the imaginary part of which only affects the higher, non-centroid, normal modes. Discarding the imaginary Liouvillian (in a rather speculative way) leads to spurious springs in the dynamics and gives RPMD.[53]

RPMD must therefore be still considered as an approach which does not represent a rigorously derived approximation to the exact quantum dynamics of the system. However, it satisfies several useful formal properties, which suggest that $k_{\mathrm{RP}}(T)$ should provide a reliable estimate of the exact quantum rate for a large class of reactions. First, RPMD rate is exact in two important model limits, namely the high-temperature and parabolic barrier.[31] Second, the distribution $\exp(-\beta_N H_N)$ gives the exact quantum Boltzmann statistics (in the limit $N \to \infty$).[26] This remarkable property is very well known and has been used for a long time to compute static properties. Third, the ring-polymer dynamics satisfies the detailed balance [*i.e.*, preserves the distribution $\exp(-\beta_N H_N)$].[26] Finally, the ring-polymer dynamics is exact in the limit $t \to 0$.[26,55] As a result of these properties, RPMD is expected to give a good estimate of the exact quantum rate coefficient for reactions that can be approximated by statistical theories. In practice, this means mainly reactions for which the rates are dominated by passage through a free-energy bottleneck, *i.e.*, the sorts of reactions that can be well approximated by TST if treated classically. In what follows, we give a detailed analysis of the ability of RPMD to treat TST-type reactions (Sections 2.2 and 2.3). However, a growing number of recent calculations[36,40,41] have found that RPMD also works well for statistical complex-forming reactions. This will be discussed further in Section 3.3.4.

## 2.2. RPMD-TST

In classical rate theory, reactions dominated by direct passage through a free-energy bottleneck can usually be well approximated by (classical) TST, which takes the instantaneous ($t \to 0_+$) flux through the bottleneck.[56,57] The same is true for such reactions in RPMD. One can approximate the RPMD rate $k_{\rm RP}(T)$ by its $t \to 0_+$ approximation, which gives the RPMD-TST rate, $k^{\#}_{\rm RP}(T)$,[27-30]

$$k^{\#}_{\rm RP}(T)Q(T) = \lim_{t \to 0_+} C_{\rm fs}(t,T) \tag{7}$$

$$= \lim_{N\to\infty} \frac{1}{(2\pi\hbar)^N} \int d\mathbf{p} \int d\mathbf{q} e^{-\beta_N H_N} \delta[f(\mathbf{q})] \dot{f}(\mathbf{q}) h[\dot{f}(\mathbf{q})] \tag{8}$$

$$= \lim_{N\to\infty} \frac{1}{2\pi\hbar\beta_N} \left(\frac{m}{2\pi\beta_N\hbar}\right)^{(N-1)/2} \int d\mathbf{q} e^{-\beta_N U_N} \sqrt{B_N(\mathbf{q})} \delta[f(\mathbf{q})] , \tag{9}$$

in which

$$B_N(\mathbf{q}) = \sum_{i=1}^{N} \left[\frac{\partial f(\mathbf{q})}{\partial q_i}\right]^2 \tag{10}$$

normalizes the flux through the dividing surface $f(\mathbf{q}) = 0$. From classical rate theory, one knows that $k^{\#}_{\rm RP}(T)$ depends strongly on $f(\mathbf{q})$, because the $t \to 0_+$ limit neglects the effects of recrossing (of trajectories that initially pass through $f(\mathbf{q}) = 0$ from reactants and products, but subsequently turn back). As a result, $k^{\#}_{\rm RP}(T) > k_{\rm RP}(T)$, which implies that the optimal $f(\mathbf{q})$ is the surface that minimizes $k^{\#}_{\rm RP}(T)$, or equivalently, maximizes the free energy at $f(\mathbf{q}) = 0$; *i.e.*, $f(\mathbf{q})$ must pass through the free-energy bottleneck. One can show that this optimal $f(\mathbf{q})$ must be invariant under cyclic permutation of the bead positions.[27] A simple $f(\mathbf{q})$ that satisfies this requirement is the centroid

$$f_{\rm centroid}(\mathbf{q}) = \frac{1}{N}\left[\sum_{i=1}^{N} q_i\right] - q^{\#} \tag{11}$$

and this important special case of RPMD-TST was suggested a long-time ago as a possible quantum TST.[13,14,58] The centroid dividing surface is often a good approximation to the optimal $f(\mathbf{q})$ for symmetric barriers, but breaks down for asymmetric barriers at low temperatures (see Section 2.3 below).

Because $k^{\#}_{\rm RP}(T)$ is a static quantity involving the distribution $\exp(-\beta_N H_N)$, and because RPMD is exact in the limit $t \to 0$, it follows that RPMD-TST must give the correct $t \to 0_+$ quantum flux through $f(\mathbf{q})$; in other words that RPMD-TST is a $t \to 0_+$ QTST. This was derived explicitly in Ref. 27, by taking the $t \to 0_+$ limit of the quantum time-correlation function in which the flux and side dividing surface have the permutationally invariant form of the optimal $f(\mathbf{q})$. It was also shown that this $t \to 0_+$ limit [*i.e.*, $k^{\#}_{\rm RP}(T)$] gives the exact quantum rate in the absence of recrossing of the dividing surface $f(\mathbf{q}) = 0$ and of surfaces orthogonal to $f(\mathbf{q})$ in path-integral space, but that $k^{\#}_{\rm RP}(T)$ does not give a strict upper bound to the exact rate (because interference effects can cause recrossing to increase the rate).[28] This last point illustrates an important limitation of RPMD-TST: if the dynamics is mostly classical (*i.e.*, does not involve any phase), $k^{\#}_{\rm RP}(T)$ gives an approximate upper bound to the exact quantum rate, allowing one to identify the optimal $f(\mathbf{q})$ as the dividing surface that maximizes the free energy; but if the dynamics is quantum (*i.e.*, involves phases) then the optimal $f(\mathbf{q})$ cannot be found. Thus, RPMD-TST is a semiclassical theory that works at temperatures sufficiently high such that dynamical phase effects are unimportant, with the majority of the quantum effects being confined to the statistics. In practice, this regime persists down to temperatures sufficiently low that quantum tunneling (mainly in the statistics) speeds up the rate by orders of magnitude; see Section 2.3 below.

For practical studies of polyatomic systems, however, it is impossible to formulate a general computational strategy of calculating the RPMD-TST rate $k^{\#}_{\rm RP}(T)$, because it is difficult to locate the optimal dividing surface $f(\mathbf{q}) = 0$. Since $k_{\rm RP}(T)$ gives a good lower bound estimate of $k^{\#}_{\rm RP}(T)$, it is better to calculate the full RPMD rate $k_{\rm RP}(T)$ instead, using adaptations of standard classical rate methods (see Section 3). Because $k_{\rm RP}(T)$ is independent of the position of $f(\mathbf{q})$, one needs to pick only a reasonable estimate of the optimal $f(\mathbf{q})$ (to ensure reasonable recrossing statistics - see Section 3) which is usually taken to be the centroid of some reaction coordinate.

### 2.3. Steepest-descent analysis of the free-energy bottleneck.

As just mentioned, RPMD-TST gives reliable predictions of reaction rate coefficients that are dominated by quantum tunneling through a free-energy bottleneck, because (except at extremely low temperatures) tunneling effects are captured mainly by the quantum Boltzmann statistics, which RPMD-TST describes correctly. We can gain insight into the nature of these statistics using a steepest-descent analysis, meaning that we locate the minimum on the ring-polymer potential $U_N(\mathbf{q})$, subject to the constraint $\delta\left[f(\mathbf{q})\right]$, and assume that the free-energy integral [over $\mathbf{q}$ in eq. (9)] can be approximated by harmonic fluctuations around this minimum. This approximation is usually sufficient to capture the essential physics, unless $V(\mathbf{q})$ is very flat.

To identify the constrained minimum on $U_N(\mathbf{q})$, we find the unconstrained first-order saddle-point on $U_N(\mathbf{q})$, then identify the dividing surface function $f(\mathbf{q})$ as the constraint which converts this saddle-point into a minimum. Differentiating $U_N(\mathbf{q})$ of eq. (3) with respect to the $q_i$, we find that stationary points on $U_N(\mathbf{q})$ must satisfy[59,60]

$$V'(q_i) = m\frac{q_{i+1} - 2q_i + q_{i-1}}{(\beta_N \hbar)^2}, \tag{12}$$

which is a discretized version of Newton's second law on the inverted potential energy surface $-V(\mathbf{q})$. The stationary points thus distribute the beads at discrete time-intervals along classical trajectories on $-V(\mathbf{q})$. Because polymers are rings, these trajectories must be periodic orbits, with period $\beta\hbar$. One can show that a saddle-point exists on $U_N(\mathbf{q})$ at every temperature, except at the *cross-over temperature* $T_{\rm c} = (k_{\rm B}\beta_{\rm c})^{-1}$, where[13]

$$\beta_c = \frac{2\pi}{\hbar\omega} \tag{13}$$

and $\omega$ is the barrier frequency. This is because $T > T_{\rm c}$ and $T < T_{\rm c}$ correspond to two different tunneling regimes, where the saddle-points are of different types; these two saddle-points coalesce at $T = T_{\rm c}$, where the steepest-descent analysis breaks down (and it is unreliable over a narrow band of temperatures on either side of $T_{\rm c}$).

At $T > T_{\rm c}$, the saddle-point corresponds to the "trivial" periodic orbit, in which all beads sit at the minimum of $-V(\mathbf{q})$ for a duration $\beta\hbar$. The unstable mode of the saddle-point is the centroid, which fixes the dividing surface $f(\mathbf{q})$ as $f_{\rm centroid}(\mathbf{q})$ of eq. (11), with $q^{\#}$ taken to be the position of the barrier maximum. The centroid represents the classical contribution to the free energy; fluctuations of the polymers around the centroid describe the tunneling through the barrier (and in a multi-dimensional system, fluctuations perpendicular to the reaction coordinate describe also the ZPE at the barrier). In the high temperature limit, these fluctuations disappear, such that $k^{\#}_{\rm RP}(T)$ becomes the classical TST counterpart. At lower temperatures, the fluctuations explore mainly the parabolic tip of the barrier, such that $k^{\#}_{\rm RP}(T)$ is given approximately by the parabolic-barrier tunneling rate. This type of tunneling, which typically speeds up the rate by less than an order of magnitude, is often referred to as "shallow" tunneling.

At $T < T_{\rm c}$ the saddle-point switches to a non-trivial periodic orbit, corresponding to a delocalized saddle-point geometry called the "instanton".[59] The instanton geometry is closed, with the beads being distributed along a line. As the temperature decreases, the instanton lengthens and, in multidimensional systems, bypasses the saddle point on $V(\mathbf{q})$, in the phenomenon known as "corner cutting". The tunneling is now "deep", in the sense that the geometry around which the polymers fluctuate (*i.e.,* the instanton) is the dominant tunneling path (rather than a classical point). Strictly speaking, the instanton is not a saddle-point, since there are $N$ instanton geometries corresponding to the $N$ permutations of the beads, which are connected by a mode whose frequency tends to be zero in the limit $N \to \infty$. At temperatures just below $T_c$, one can constrain approximately the unstable modes of all $N$ of the instantons using the centroid dividing surface, and this approximation remains good for symmetric barriers down to about $T_c$ / 2. However, for asymmetric barriers, the centroid gives a poor approximation at low temperatures, and must be replaced by a more general permutationally-invariant dividing surface.

One can also use the steepest-descent analysis to estimate the effects of recrossing on the RPMD-TST rate coefficient at low temperatures, by comparing with the predictions of instanton theory, obtained by making the Van Vleck-Gutzwiller approximation to the Miller-Schwarz-Tromp expression for the exact quantum rate coefficient.[61] One finds that $k_{\mathrm{RP}}^{\#}(T)$ remains a good approximation of the instanton rate down to about $T_c/2$, which is underestimated if the barrier is reasonably symmetric, and overestimated if the barrier is significantly asymmetric. These predictions, based on a one-dimensional steepest-descent analysis, turned out to work very well for realistic chemical reactions (see Section 3). However, more work is needed to understand better how RPMD rate theory behaves at extremely low temperatures.

Note that the analysis presented above explains why short-time limit of RPMD rate theory, $k_{\mathrm{RP}}^{\#}(T)$, is expected to be very accurate in certain cases. However, as discussed above, the connection between long- and short-time limits, as it presently stands, is derived in a rather heuristic way and therefore practical validation using realistic systems is required for assessing RPMD rate theory. This will be discussed in the following Section 3.

## 3. COMPUTATIONAL PROCEDURE AND PRACTICAL APPLICATIONS

### 3.1. RPMDrate code

Thermal rate coefficients for bimolecular chemical reactions in the gas phase (A + B → products) are calculated by combining the RPMD rate theory described in Section 2 with the Bennett-Chandler method[62,63] and the formalism of two dividing surfaces, which is in many ways similar to the one originally proposed by Takayanagi and Miller for the path integral evaluation of the quantum instanton (QI) rate coefficients.[64] This approach has been discussed in detail in Refs. 33 and 51, so only the key working expressions are presented here.

The computational procedure begins by introducing two dividing surfaces for a given reaction of the reactant species A and B. The first one is located in the reactant asymptotic valley

$$s_0(\bar{\mathbf{q}}) = R_\infty - \left|\bar{\mathbf{R}}\right|, \tag{14}$$

where $R_\infty$ is an adjustable parameter of distance between two centers of mass of A and B that is chosen to be sufficiently large that the interaction potential vanishes, $\left|\bar{\mathbf{R}}\right|$ is the centroid of the vector that connects the two centers of mass of the two reactants,

$$\bar{\mathbf{R}} = \frac{1}{m_{\mathrm{B}}}\sum_{i_{\mathrm{B}}=1}^{N_{\mathrm{B}}} m_{i_{\mathrm{B}}}\bar{\mathbf{q}}_{i_{\mathrm{B}}} - \frac{1}{m_{\mathrm{A}}}\sum_{i_{\mathrm{A}}=1}^{N_{\mathrm{A}}} m_{i_{\mathrm{A}}}\bar{\mathbf{q}}_{i_{\mathrm{A}}}, \tag{15}$$

with $m_{\mathrm{X}} = \sum_{i_{\mathrm{X}}=1}^{N_{\mathrm{X}}} m_{i_{\mathrm{X}}}$ and $\bar{\mathbf{q}}_{i_{\mathrm{X}}} = \frac{1}{N}\sum_{j=1}^{n_b} \mathbf{q}_{i_{\mathrm{X}}}^{(j)}$.

The second dividing surface is located in the "transition-state" region, which can be chosen at the saddle point for thermally activated reactions or near the complex-formation minimum for complex-forming reactions, and is defined in terms of the distances of bonds that break and form as

$$s_1(\bar{\mathbf{q}}) = \min\left\{s_{1,1}(\bar{\mathbf{q}}),\ldots,s_{1,N_{\mathrm{bonds}}}(\bar{\mathbf{q}})\right\}, \tag{16}$$

where $N_{\mathrm{bonds}}$ is the number of relevant combinations of breaking and forming bonds and

$$s_{1,k}(\bar{\mathbf{q}}) = \max\left\{s_{1,k}^{(1)}(\bar{\mathbf{q}}),\ldots,s_{1,k}^{(N_{\mathrm{channel}})}(\bar{\mathbf{q}})\right\}, \tag{17}$$

where $N_{\mathrm{channel}}$ is the number of equivalent product channels and

$$s_{1,k}^{(l)}(\bar{\mathbf{q}}) = \left(\left|\bar{\mathbf{q}}_{12,k}^{(l)}\right| - q_{12,k}^{\#(l)}\right) - \left(\left|\bar{\mathbf{q}}_{23,k}^{(l)}\right| - q_{23,k}^{\#(l)}\right), \tag{18}$$

is the dividing surface for each individual combination (one bond breaks between atoms 1 and 2 and one bond forms between atoms 2 and 3). In eq. (18), $\bar{\mathbf{q}}_{ij}$ denotes the vector connecting the centers of mass of atoms *i* and *j*, while $q_{ij}^{\#}$ is the corresponding distance at the fixed configuration used to define the "transition-state" coordinate. Such definition of $s_1(\bar{\mathbf{q}})$ in eqs. (16)-(18) is rather general and can be applied to very complex polyatomic chemical reactions with several bonds that can simultaneously break and form. However, to our knowledge, only reactions with a single event of bond breakage/formation have been studied so far.

The next step is to generate the reaction coordinate $\xi$ that connects the reactants region with the "transition-state" region using the dividing surfaces

$$\xi(\bar{\mathbf{q}}) = \frac{s_0(\bar{\mathbf{q}})}{s_0(\bar{\mathbf{q}}) - s_1(\bar{\mathbf{q}})}, \tag{19}$$

Here $\xi \to 0$ as $s_0 \to 0$ and $\xi \to 1$ as $s_1 \to 0$.

The final step is to represent the RPMD rate coefficient in the Bennett-Chandler form[62,63]

$$k_{\mathrm{RP}}(T) = k_{\mathrm{cd-QTST}}(T;\xi^{\#})\kappa(t \to t_p;\xi^{\#}). \tag{20}$$

Here $k_{\mathrm{cd-QTST}}(T;\xi^{\#})$ is the centroid-density quantum transition-state theory (cd-QTST) and $\kappa(t \to t_p;\xi^{\#})$ is the long-time limit of the time-dependent ring-polymer transmission coefficient.

The first factor in eq. (20) is evaluated by using the well-established formalism of the centroid potential of mean force (or free energy) along $\xi$

$$k_{\mathrm{cd\text{-}QTST}}(T) = 4\pi R_{\infty}^2 \left(\frac{m_{\mathrm{A}} + m_{\mathrm{B}}}{2\pi\beta m_{\mathrm{A}} m_{\mathrm{B}}}\right)^{1/2} e^{-\beta\left[W(\xi^{\#}) - W(0)\right]}. \tag{21}$$

The free-energy difference in eq. (21) can be calculated by umbrella integration.[65,66] Note that the ring-polymer Hamiltonian used to calculate this value differs from the original one[33]

$$\tilde{H}_n(\mathbf{p},\mathbf{q}) = H_n(\mathbf{p},\mathbf{q}) - \frac{1}{\beta_n}\ln f_s(\bar{\mathbf{q}}), \quad (22)$$

with

$$f_s(\bar{\mathbf{q}}) = \left\{ \sum_{i=1}^{N_A+N_B} \frac{1}{2\pi\beta m_i} \left| \frac{\partial \xi(\bar{\mathbf{q}})}{\partial \bar{\mathbf{q}}_i} \right|^2 \right\}^{1/2}. \quad (23)$$

The second factor in eq. (20), $\kappa(t \to t_p;\xi^{\#})$ (ring-polymer recrossing factor or transmission coefficient), represents a dynamical correction to $k_{\mathrm{cd-QTST}}(T;\xi^{\#})$ that accounts for recrossing of the transition-state dividing surface at $t \to t_p$, where $t_p$ is a "plateau" time that ensures that the resulting RPMD rate coefficient $k_{RP}$(T) will be independent of our choice of $s_0(\bar{\mathbf{q}})$ , $s_1(\bar{\mathbf{q}})$, and thus $\xi(\bar{\mathbf{q}})$. It is calculated using a combination of a long constrained at $\xi^{\#}$ "parent" ring-polymer trajectory and a series of shorter unconstrained "child" ring-polymer trajectories.[51] The optimal value of $\xi^{\#}$ is determined from the free-energy profile and is selected at the maximum value of $W(\xi)$ in order to reduce the number of recrossings of the dividing surface and to ensure faster convergence of $\kappa(t \to t_p;\xi^{\#})$. Note that in RPMD the choice of the dividing surface in the transition-state region is unimportant as the static factor ($k_{\mathrm{cd-QTST}}(T;\xi^{\#})$) is counterbalanced by the dynamic factor ($\kappa(t \to t_p;\xi^{\#})$).

The computational procedure described above was implemented in the RPMDrate code developed by one of us (Y.V.S.). Most of the results discussed in the present review were generated using this code. A complete description of the code and typical input parameters can be found in Ref. 51. In general, the computational strategy mainly depends on the type of chemical reaction under investigation. For chemical reactions with a barrier, the most demanding computation is the construction of the free-energy profile that may require "more intense" (narrow window spacing) umbrella sampling near the transition state,[34] while the subsequent calculation of the transmission coefficient is much faster due to a small value of $t_p$. For complex-forming chemical reactions, the free-energy profile is rather flat in the long-range region and therefore umbrella integration requires less computation time (wider umbrella windows for biased sampling), while the main computation expense is due to recrossing dynamics ($\kappa(t)$) in the complex-formation region which results in significantly higher values of $t_p$.[40,41] It is therefore important to find an optimal balance between the input parameters for calculating the two factors in eq. (20) depending on the type of the chemical reaction under study.

### 3.2. Number of Beads

Convergence of RPMD with respect to the total number of replicas, or beads ($n_b$), is another important aspect of its practical applications, because the total number of gradient calls ($\partial V / \partial q$) scales linearly with $n_b$ in any RPMD simulation. This means that RPMD calculations are roughly $n_b$ times slower that the purely classical ones with $n_b$ =1. It is well established that the number of ring-polymer beads for converged path integral (PI) calculations (either PI molecular dynamics (PIMD) or RPMD) should satisfy the following relationship[34]

$$n_{\mathrm{b}} > n_{\mathrm{min}} \equiv \frac{\hbar\omega_{\mathrm{max}}}{k_{\mathrm{B}}T}, \qquad (24)$$

where $\omega_{\mathrm{max}}$ is the largest frequency of the system. From eq. (24) it is clear that the convergence with respect to $n_{\mathrm{b}}$ depends on both the system and the temperature. At high temperatures (*e.g.*, $T \geq 1000$ K, relevant to combustion chemistry), RPMD rate coefficients can be converged very quickly with only a small number of beads ($n_{\mathrm{b}}$ = 4-8),[34,35] which satisfies the criterion in eq. (24). Exceptions are only found in cases of very light atom transfers, such as Mu-transfer in the D + HMu → DMu + H reaction, which may require a high number of beads even at high temperatures.[23] At intermediate temperatures, the total number of beads is usually within the range of 16-64. At lowest temperatures considered in most of the cases (near 200-300 K), this number increases up to 64-192 – within a traditional factor of two orders of magnitude of computational costs increase for PI simulations compare to classical ones near the room temperature. The lowest temperature considered so far was $T$ = 50 K for the C($^1D$) + $H_2$ reaction ($n_{\mathrm{b}}$ = 256)[36] and the highest number of beads $n_{\mathrm{b}}$ = 512 was used to converge the RPMD rates of the D + MuH → DMu + H reaction ($T$ = 150-1000 K).[23] Note that the total number of beads was not fully optimized for many systems with analytical PESs due to very fast evaluations of $\partial V / \partial q$. However, these calculations were still much faster than the exact QM ones for which computation time increases exponentially with the number of atoms in the system prohibiting any rigorous calculations for polyatomic dynamics (except symmetric X + $CH_4$ systems)[8-10,67,68] with the present computer power. Note that the convergence of RPMD with respect to $n_{\mathrm{b}}$ is in many ways very similar to the one for path integral evaluation of the quantum instanton thermal rates[64] except the fact that in the RPMD case the path integrals are evaluated over the whole space between the reactants and the products (which can affect $\omega_{\mathrm{max}}$).

It is also interesting to note several attempts to reduce the number of beads in the RPMD calculations and thus reduce the cost of evaluating the forces. One approach is to use special contraction schemes for representing the interaction potential which would allow efficient combination of high- and low-level electronic structure theory and enable lower values of $n_{\mathrm{b}}$ for certain contributions to the total PES.[69-73] However, these approaches have been mainly focused on model and *ab initio* (MP2/DFT) simulations of several water systems (condensed phase, dimers, cations). Their general applicability to polyatomic chemical reactions requires further investigation. Another approach is to use different values of $n_{\mathrm{b}}$ for converging $k_{\mathrm{cd\text{-}QTST}}$ and $\kappa$(t) in eq. (21). In principle, this approach is based on a similar idea of splitting the interaction potential since, unlike $\kappa(t)$, the ring-polymer trajectories used to evaluate of $k_{\mathrm{cd\text{-}QTST}}$ do not "visit" the product side. Recent study of the H + $CH_4$ reaction demonstrated a possibility to converge the RPMD thermal rates with a considerably reduced number of replicas,[74] although further testing of this approach is also required.

## 3.3. Bimolecular Chemical Reaction Rate Coefficients from RPMD

Previous RPMD studies of chemical kinetics can be divided into three categories: (1) method's accuracy assessment using small prototype chemical reactions (one-dimensional model barriers and gas-phase systems) with analytical PESs, *i.e.*, RPMD thermal rates versus the corresponding rigorous QM ones;[23,31,33,34,37-42] (2) comparison between various dynamical approximations (including RPMD) and experiment for polyatomic gas-phase chemical reactions with more realistic PESs, for which QM results are either scarce or unavailable;[24,35,36,43-46,74-76] (3) RPMD mechanistic studies of complex condensed-phase chemical dynamics.[47-50] Although the results from the third category are very important from the practical point of view since they demonstrated that RPMD can be successfully applied to very complex systems, such as enzyme catalysis or metal surface dynamics, the present review is focused on the first two categories - the comparative analysis performed in these studies allowed us to formulate the main conclusions regarding the method's accuracy and reliability for calculating thermal rate coefficients.

Tables 1 and 2 summarize the results generated within the first category (RPMD rate coefficients and their deviations from the rigorous QM values). Table 1 includes the following types of thermally activated (with a reaction barrier) chemical reactions: 1D Eckart model; atom-diatom chemical reactions; prototype polyatomic chemical reactions, *i.e.*, hydrogen abstraction from methane by the hydrogen, chlorine or oxygen atom. Table 2 includes prototypical atom-diatom complex-forming reactions in which the atom is electronically excited. The results from the second category are also presented in our further discussion, which is divided into several thematic areas for the reasons that will become clear on reading.

*3.3.1. Quantum Mechanical Tunneling. Reaction Symmetry Assignment*

Due to a complex multidimensional nature and the dominant role at low temperatures, quantum tunneling represents, perhaps, the most challenging issue in a proper treatment of quantum mechanical nuclear motion by various quantum dynamics approximations. The crossover temperature, $T_c$, introduced in the previous Section 2.3 (see eq. (13)) is used to characterize the importance of quantum tunneling for a given reaction. Though for certain PES profiles significant tunneling is possible even above $T_c$,[77] the crossover temperature often serves as a preliminary indicator - temperatures below $T_c$ are usually referred as to the "deep-tunneling" regime where tunneling plays a dominant role. As discussed in Section 2.3, in this range of temperatures the Boltzmann matrix is dominated by fluctuations around the delocalized "instanton" path, which is a periodic orbit on the inverted potential well. Instanton theory resulted in several practical approximate methods for rate calculations developed in the past,[15,16,78-85] which aim to treat quantum tunneling particularly accurately. Using the fact that a finite difference approximation to the instanton trajectory appears as a saddle point on the ring-polymer PES, Richardson and Althorpe have established a connection between RPMD and one of these theories, namely, the uniform *ImF* version.[59] Therefore, the RPMD rate coefficient was *a priori* expected not to fail in the quantum tunneling regime.

Table 1 shows that RPMD provides a consistent and reliable behavior at low temperatures across a wide range of system dimensionalities. For all the systems with a barrier, below $T_c$, RPMD systematically underestimates (overestimates) the rates for symmetric (asymmetric). Note that it was found that in order to assign the reaction symmetry for realistic systems it is important to take into account the ZPE difference between the reactant and product sides (which can be estimated from the vibrationally adiabatic reaction profile).[39] With one exception, the D + MuH reaction, which will be discussed later, the lowest temperatures considered in the RPMD studies were around $T_c/2$. The percentage errors given in Table 1 indicate that RPMD is usually within a factor of 2-3 at the lowest temperatures, as determined by comparison to rigorous QM results available for these systems. Note that the most challenging systems are those with a transfer of light atoms from/between much heavier partners, such as Cl + HCl, D + MuH, and H + $CH_4$. Nevertheless, the accuracy of RPMD is indeed remarkable. Figure 1 shows that RPMD provides excellent agreement with exact QM results for the prototype polyatomic reaction H + $CH_4$, significantly better than those obtained using centroid-density quantum TST (cd-QTST and QTST in Figure 1), quantum instanton theory (QI in Figure 1) or canonical variational transition-state theory with microcanonically optimized multidimensional tunneling correction (CVT/$\mu$OMT in Figure 1). One of the main issues of TST-based methods is their sensitivity to the choice of the transition-state dividing surface. Proper identification of this dividing surface becomes increasingly difficult as the dimensionality of the problem rises due to the multidimensional nature of tunneling at low temperatures. As a result, TST-based methods are often less accurate in higher dimensionalities. Moreover, elaborate quantum implementations of TST have similar sensitivity to the definition of the dividing surface. As a result, they do not necessarily provide improved accuracy when compared to less sophisticated methods, and therefore are not guaranteed to provide a predictable level of error when applied to higher-dimensionality systems. This has been observed when shifting from 1D models to realistic gas-phase systems and even to multidimensional models.[86] The higher (predictable) level of accuracy, due to the rigorous independence of RPMD from the choice of transition-state dividing surface, is one of the most important and attractive features of RPMD rate theory when compared to TST-based methods.

Another and, perhaps, the most representative comparative study of the QM tunneling effect was performed for the D + HMu → DMu + H reaction.[23] Despite its apparent simplicity (atom-diatom), this reaction can be considered as the most challenging stress test for dynamical methods in the deep tunneling regime. The transfer of the Mu atom, which is a light isotope of hydrogen and has a mass of just 0.11398 u, between two significantly heavier hydrogen isotopes gives rise to a dominant tunneling effect and a very high crossover temperature ($T_c$ = 860 K). This reaction exhibits a very small skewing angle and gives rise to a deep well in the ZPE-corrected PES near the saddle point (more detail can be found in Ref. 23). This reaction is also symmetric in the deep-tunneling regime and, as a result, RPMD is expected to underestimate the thermal rate coefficient. Figure 2 presents a detailed comparison of the performance of different methods as the ratio of the *i*CVT/$\mu$OMT (where "*i*" stands for an improved treatment of the energetic reaction threshold region), semiclassical instanton (SCI), and RPMD results and the exact QM. It shows that RPMD provided better in overall accuracy in the whole temperature range and more stable results with deviations from the exact QM results that barely change with temperature and within the expected factors. Both representative TST-based methods chosen for this study demonstrated irregular behavior with one method overestimating the rate coefficient (SCI) and another underestimating it (*i*CVT). Alhough, as discussed previously, RPMD is related to SCI, but it has the advantage of being free of any adjustable empirical parameters (usually used to "tune" the transition-state dividing surface and the corresponding reaction path) and this is likely the origin of its better reliability for this stress test. TST-based methods, including those used in this study, contain certain approximations to the tunneling path. In particular, the $\mu$OMT correction to *i*CVT assumes that that tunneling is initiated by vibrational motions perpendicular to the reaction coordinate rather than by the motion along the reaction coordinate and it uses the straight-line path approximation in evaluating the largest tunneling probability. The SCI uses the harmonic approximation perpendicular to the instanton coordinate. In addition, both approximations employ non-recrossing assumption, likewise any other TST-based methods.[23] Because of the complex multidimensional nature of quantum tunneling, this example, that constitutes a true stress test, demonstrates that such assumptions can fail even for triatomic systems in the deep-tunneling regime.

The quality of RPMD in calculating rate coefficients for reactions involving polyatomic molecules such as methane is also satisfactory, as shown in Table 1. The deviations from approximate quantum dynamical results either above or below $T_c$ are comparable to those found in atom-diatom systems. The good agreement with quantum dynamical results in the deep-tunneling regime is particularly impressive as multidimensional tunneling is very hard to describe. In addition, the RPMD results for the Mu + $CH_4$ reaction differ significantly from the CVT/$\mu$OMT counterparts, presumably due to the errors of latter in treating the tunneling or in ignoring anharmonicity.[42]

### *3.3.2. Zero-Point Energy Effect*

Another important quantum mechanical effect of nuclear motion is the ZPE. As an example of how well RPMD captures ZPE effects, the Mu + $H_2$ → MuH + H reaction demonstrates a notoriously large difference in ZPE between reactants and products, which results in an inverse kinetic isotope effect observed theoretically and confirmed experimentally (See Figure 4 in Ref. 38). Table 1 and Figure 3 show that throughout the temperature range 200-1000 K, the RPMD rate coefficient for this reaction is very close to the exact quantum mechanical result, within 15 % error at the lowest temperature considered in this study. Similar observations are made for the H/D/Mu + $CH_4$ reactions.[42]

A related issue is the accurate treatment of the ZPE effect near saddle points. Improper treatment of anharmonicity of high-frequency modes near transition states can lead to significant errors in conventional TST calculations (VTST), such as rate coefficient values and temperature dependence.[22] One would expect that anharmonicity plays a significant role at high temperatures, where higher vibrational levels are more populated. As mentioned above, the D + MuH reaction represents an example with a very complex structure of the ZPE near transition state, but its effect for this reaction is dissembled by a dominant role of tunneling and very high crossover temperature. More distinct examples include our recent study of the O + $CH_4$ → $CH_3$ + OH and Mu + $CH_4$ → $CH_3$ + HMu reactions.[21,42] A significant disagreement between variational transition-state theory with multidimensional tunneling corrections (VTST/MT) calculations and experiment was found in the classical high temperature limit for the former reaction. Surprisingly, VTST rates were lower than the experimental ones by a factor of two while the RPMD results matched perfectly the experimental data. Note that RPMD is exact in the high-temperature limit. Therefore, this discrepancy has been attributed to the harmonic separable-modes approximation used in the VTST calculations to compute the corresponding partition functions. As discussed above, the large difference between the RPMD and CVT/$\mu$OMT results for the latter reaction might also be due to the harmonic treatment in the TST approach. Though recent developments in the conventional TST allow one to include approximately anharmonic effects,[22] it is important to emphasize that such improvements are often "one-sided" in the sense that they are focused only on one source of error in TST calculations (*e.g.*, anharmonicity, multiple reaction paths, coupling of different types of motion, or tunneling contributions, *etc.*). As a result, such modifications can lead to even worse agreement with experiment than earlier versions of TST, because some error cancellation disappears. Unlike TST, RPMD is immune to such issues as this method treats all degrees of freedom on an equal footing, accurately takes into account the ZPE effect along the reaction coordinate and demonstrates predictable and consistent treatment of quantum tunneling effect.

### *3.3.3. Kinetic Isotope Effects*

Kinetic isotope effects (KIE) serve as another way to assess accuracy since it is widely used in experiments to detect quantum mechanical effects. For simple atom-diatom systems, such as X + $H_2$ (X = H, D, Mu, He$\mu$), RPMD reproduced very accurately the variation of the exact QM isotope effects with temperature for all of the various isotopic variants.[39] For more complex systems studied using RPMD, such as hydrogen abstraction from methane by atoms (H,[42] O,[34] and Cl[35]) or by diatomic radical OH,[24,45] such comparison with the exact KIEs is not always possible due to the lack of QM calculations for the other isotopes (six-atom reactions) or to the complete absence of any rigorous QM results (the seven-atom reaction). In those cases, the comparison with experiment represents an alternative. As a ratio of rate coefficients, the KIE depends weakly on the height of the classical reaction barrier, and therefore provides a useful means for comparison to experiment even when the PES is less accurate. While either scarce or scattered experimental data are available for six-atom reactions, the seven-atom reactions OH + $CH_4$ → $CH_3$ + $H_2O$ and OH + $CD_4$ → $CD_3$ + HDO provide excellent opportunity to test the dynamics approaches. These KIEs have received considerable experimental attention and reliable experimental results were obtained in independent measurements. Table 3 shows that although there have been several TST calculations of the KIE, generally they are not consistent with one another or the experimental results. CVT and QI methods overpredict the KIE and the VTST/CVT and CUS/μOMT methods underestimate the KIE. The oldest and least sophisticated CVT calculation is the most accurate, though this accuracy, as discussed above, is likely to be the result of error cancellation. At low temperatures, where tunneling plays very important role, the QI method exhibits poor performance compare to less sophisticated cd-QTST, though both calculations were performed using the same PES. By contrast, in the previous studies of the 1D Eckart model and gas-phase atom-diatom reactions, the QI method often provides equal or more accurate results than those of CVT, cd-QTST, or RPMD rate theory. The reason for such inconsistent, unpredictable performance of the TST-based methods is again their sensitivity to the choice of transition-state dividing surface, which becomes an issue with increasing the reaction dimensionality. At the same time, RPMD demonstrates a consistent and predictable behavior for this seven-atom reaction. At low temperatures, it can be expected that RPMD will overestimate the rate coefficient for the asymmetric reaction in the numerator of the KIE ($CH_4$) by a factor of 2. Since the reaction in the denominator (H + $CD_4$) has a smaller tunneling contribution than the numerator (due to deuterium transfer), RPMD could be very accurate. As a result, we anticipated that the RPMD rate theory will overestimate the KIE by a factor of 2 at low temperatures – in agreement with the results in Table 3. At high temperatures, RPMD is also significantly more accurate than the other TST-based methods, where recrossing dynamics becomes very important.

Another stringent test is the $^{12}C/^{13}C$ KIEs for this seven-atom reaction, which are strongly dependent on the dynamical method used to treat the torsional anharmonicity of the lowest-frequency vibrational mode and are less dependent on the PES used.[45] Table 3 shows that RPMD is more accurate than TST and underestimates the experimental value by a very small amount of less than 1%, which is even better than the statistical uncertainty in the RPMD calculations or the experimental uncertainty. Such excellent agreement is due to the fact that RPMD treats all degrees of freedom on an equal footing and, as a result, is truly full dimensional.

### *3.3.4. Insertion Reactions*

So far our discussion has been focused on thermally activated reactions. Barrierless complex-forming reactions represent another class of chemical reactions that plays an important role in combustion, atmospheres, and cold interstellar media.[12,36,40,41] Due to the formation of intermediate collision complex(es), rigorous quantum dynamical (QD) studies of such reactions are much more computationally demanding and therefore their rigorous QM treatment lag significantly behind those for reactions dominated by a significant barrier.[12] There are two obvious difficulties associated with a barrierless reaction profile – (1) it is rather difficult to locate a free-energy barrier (which is usually relatively small and is in the asymptotic regions) and (2) the presence of long-lived intermediates can result in significant recrossing effects, especially when the rate for energy randomization is slow. Because of its independence of the dividing surface, inclusion of recrossing effects, and accurate treatment of the ZPE along the reaction coordinate, RPMD was again *a priori* expected to provide an accurate alternative to conventional TST and QCT approaches.

Another potential issue could be the quantum mechanical resonances due to the reaction intermediate, which have been observed in numerous QD studies of triatomic insertion systems.[12] As discussed above, RPMD is capable of capturing only short-time quantum dynamics effects (below the thermal time $\sim \beta\hbar$)[55] but it does not include long-time effects such as interference. Luckily, these previous QD studies demonstrated that such resonances (both short- and long-lived), though show up in detailed magnitudes as integral and differential cross sections, do not affect thermally averaged rate coefficients for the prototype triatomic insertion reactions.[12] Certainly, assessing accuracy of RPMD for reactions with more pronounced resonant structure, such as F + $CH_4$,[87] is desirable in future.

Table 2 shows that RPMD provides a remarkable level of accuracy for this class of chemical reactions. For all four prototypical triatomic systems (X + $H_2$; X = N($^2$D), O($^1$D), C($^1$D), S($^1$D)), RPMD values are in a very good agreement with the QD results within an error of ~ ± 5-10%. Note that this deviation is in fact within the statistical sampling uncertainties in the RPMD calculations. However, it may also be due in part to approximations made in the QD calculations, such as cutoff of higher rotational energy levels and a quantum statistical model. While RPMD and QD calculations agree very closely with each other, the accordance with the experimental measurements is not so good (*e.g.*, see Figure 4). Most likely the discrepancy between theory and experiment is due to deficiencies in the PES used in these studies. Our recent RPMD study of the C($^1$D) + $H_2$ also demonstrated that inclusion of excited PESs (such as singlet PES $^1A''$ that correlates with the same ground-state products) significantly improves the agreement with experiment.[36] Table 2 also suggests less sensitivity of RPMD to the energetics of insertion-type chemical reactions compared to the energetics of the thermally activated ones. Within the reaction symmetry assignment used for thermally activated chemical reactions,[39] the C($^1D$) +$H_2$ and S($^1D$) +$H_2$ reactions could be classified as "effectively symmetric".[41] As indicated above, RPMD systematically underestimates the low-temperature rates for thermally activated symmetric reaction. Table 2 shows that this rule, which was initially developed and validated for reactions with significant barriers, does not work for insertion reactions. Clearly, further accuracy assessment of RPMD for barrierless complex-forming reactions is required, although it is limited by very scarce QD results for reactions of this type. Nevertheless, the level of accuracy RPMD provided to prototypical systems suggests that it has the potential to be used for polyatomic reactions involving reaction intermediates.

## 4. DISCUSSION AND PERSPECTS ON FUTURE DIRECTIONS

The computational procedure for calculating ring-polymer molecular dynamics (RPMD) bimolecular chemical reaction rate coefficients presented in Section 3.1 was proposed in 2009-2011[33,37] and implemented in the general computer program RPMDrate in 2012-2013.[51] In subsequent years (2012-2016), the RPMDrate code was actively applied to numerous prototypical atom-diatom and polyatomic reactions with analytical potential energy surfaces (PESs),[21,23,24,34-36,38-46,50,51,74-76] as discussed in Section 3.3. These preliminary applications of RPMD rate theory can be considered as “method assessments” as the main focus of these studies was on comparison with other dynamics approximations. However, they have demonstrated that RPMD provides systematic and consistent efficiency across a wide range of system dimensionalities. For thermally activated reactions, these results demonstrated that the performance of RPMD rate theory contrasts with that of conventional methods based on the transition-state theory (TST), which are sensitive to the choice of the dividing surface. Despite the connection between RPMD short-time limit and quantum TST (see Section 2.2), its long-time limit is rigorously independent of this dividing surface.[32] The proper identification of this parameter becomes increasingly difficult as the dimensionality of the problem increases due to the multidimensional nature of tunneling at low temperatures and, sometimes, large amount of recrossings of the dividing surface at high-temperatures. As a result, our comparative analysis demonstrated that TST methods could be unreliable in higher dimensionalities. Tailoring the dividing surface for each reaction can significantly improve the accuracy of TST calculations, but it requires manual work and special expertise, and for some systems even the optimal dividing surface would still not eliminate recrossing completely. Elaborated quantum/semiclassical implementations of TST have similar sensitivities to the definition of the dividing surface, and our previous studies showed that they do not necessarily provide improved accuracy when compared to less sophisticated methods, and therefore are not guaranteed to provide a predictable level of error when applied to higher-dimensionality systems. For barrierless prototypical complex-forming reactions mediated by a deep well, RPMD has demonstrated a remarkable agreement with quantum mechanical results thus firmly establishing the validity of this method. It can be expected that the RPMD method will find wide applications in both activated and complex-forming reactions of larger sizes. The success of the RPMD rate theory for polyatomic gas phase reactions also does a lot to legitimize the previous condensed phase applications,[47,50] for which there are no exact benchmark calculations for comparison.

Regarding further development of the RPMD rate theory, it would be of considerable benefit to improve the existing computational strategy. The most topical issue is the development of efficient ways to couple RPMD with *ab initio* electronic structure evaluations in order for RPMD to become a generally useful tool. In principle, the PES can also be calculated “on-the-fly”, but even with the most advanced supercomputers it is an extremely computationally intensive procedure and is generally limited to fairly short propagation times and relatively low levels of electronic structure theory. The traditional approach for fitting PESs based on a large number of *ab initio* points has recently received a huge boost from advances in both electronic structure theory and fitting techniques.[4,5] These developments have already benefitted RPMDrate calculations of realistic (non-prototype) reactions.[46,74-76,88]

The RPMDrate code was originally developed for the simulation of dynamical processes which can be described by a single PES.[51] There are many chemical events, called non-Born-Oppenheimer reactions or electronically nonadiabatic reactions, which cannot be adequately described within this assumption. Nonadiabatic electronic transitions between electronic states play a pivotal role in numerous chemical rate processes of current interest, including electrochemical reactions, ion-molecule reactions, proton-coupled electron transfer, photoisomerization as well as many bimolecular reactions initiated in an electronically excited state and such large systems as amino acids and in transition metal centers in enzymes.[89-91] There are ongoing attempts to extend RPMD to nonadiabatic processes,[92-96] suggesting that such calculations might be feasible in the near future. However, due to a lack of benchmark prototypical systems and comparative analysis with rigorous QM results, many questions still remain to be answered before nonadiabatic generalization of RPMD becomes practically extended.

Recent studies also suggest that RPMD might well be just a first step in the emergence of a new reaction rate theory. An explicit connection between RPMD and another approximation based on the quantum Boltzmann real time-correlation functions and the path integrals (PI) formulation called "Matsubara dynamics" has been derived.[52-54] Being more accurate, "Matsubara dynamics" is too computationally expensive to be applied to real systems. However, the explicit formulae for the terms that are left out when switching to RPMD suggest that further theory development may lead to improvements in these methods. It was also shown that it is possible to apply an internal mode thermostat to RPMD without altering any of its established properties.[97] The resulting "thermostatted" version of RPMD rate theory provides more flexibility and accuracy control as various thermostatting schemes can be implemented.[98] An important point is that methods based on path integral simulations, including RPMD, are basically classical molecular dynamics methods in an extended phase space, and therefore possess favorable scaling with the size of the system and could potentially be used to calculate rate coefficients for systems containing many hundreds of atoms. There is clearly a tremendous amount of leeway in the construction of extended phase space approximations to quantum dynamics.

## ACKNOWLEDGEMENTS

We would like to thank our colleagues, in particular David E. Manolopoulos, William H. Green, Joaquin Espinosa-Garcia, Yongle Li, Ricardo Pérez de Tudela for long-standing collaboration and Stuart C. Althorpe, Timothy J. H. Hele and Jeremy O. Richardson for useful discussions. Y.V.S. thanks the European Regional Development Fund and the Republic of Cyprus for support through the Research Promotion Foundation (Project No. Cy-Tera ΝΕΑ ΥΠΟΔΟΜΗ/ΣΤΡΑΤΗ/0308/31). Y.V.S. also acknowledges funding from the Newton International Alumni Scheme from the Royal Society. F.J.A. has been funded by the Spanish Ministry of Science and Innovation under grant CSD2009-00038 and by the MINECO of Spain under grants CTQ2012-37404, CTQ-2015-65033-P. H.G. is grateful for the generous support from the United States Department of Energy (Grant numbers DE-FG02-05ER15694 and DE-SC0015997).

**Table 1. Percentage Deviations of the RPMD Thermal Rate Coefficients for Various Prototype Thermally Activated Chemical Reactions from the More Accurate Quantum Mechanical Results**

| System | % Error[a] | | | | | Reference |
|---|---|---|---|---|---|---|
| 1D models | $T < T_c$ | | $T > T_c$ | | $T_c$(K)[b] | |
| | From | To | From | To | | |
| symmetric Eckart barrier | -45 (125.5 K) | -23 (188 K) | -15 (251 K) | -10 (377 K) | 239 | ref 59 |
| asymmetric Eckart barrier | +44 (*β*=*12*) | +8 (*β*=8) | 0 (*β* = *6*) | 0 (*β* =2) | $\beta_c = 2\pi$[c] | ref 32 |
| **Thermally activated atom-diatom chemical reactions** | | | | | | |
| - energetically symmetric | | | | | | |
| $H + H_2 \rightarrow H_2 + H$ | -42 (200 K) | -33 (300 K) | -30 (400 K) | -15 (1000 K) | 345 | ref 39 |
| $D + H_2 \rightarrow DH + H$ | –38 (200 K) | - | -32 (300 K) | -15 (1000 K) | 245 | ref 39 |
| $He\mu + H_2 \rightarrow He\mu H + H$ | - | - | -37 (200 K) | -14 (1000 K) | 170 | ref 39 |
| $D + MuH \rightarrow DMu + H$ | -81 (150 K) | -57 (800 K) | -53 (900 K) | -50 (1000 K) | 860 | ref 23 |
| $Cl + HCl \rightarrow ClH + Cl$ | -70 (200 K) | -65 (300 K) | -63 (400 K) | -59 (500 K) | 320 | ref 37 |
| - energetically asymmetric | | | | | | |
| $Mu + H_2 \rightarrow MuH + H$ | +15 (200 K) | +1 (400 K) | -6 (500 K) | -2 (1000 K) | 409 | ref 39 |
| $F + H_2 \rightarrow FH + H$ | +37 (200 K) | - | +10 (300 K) | +4 (500 K) | 264 | ref 37 |
| **Thermally activated X + $CH_4$ chemical reactions** | | | | | | |
| - energetically asymmetric | | | | | | |
| $H + CH_4 \rightarrow CH_3 + H_2$ | +92 (225 K) | - | +68 (300 K) | +40 (400 K) | 296 | ref 33 |
| $D + CH_4 \rightarrow CH_3 + HD$ | +13 (300 K) | - | - | - | 341 | ref 42 |
| $O + CH_4 \rightarrow CH_3 + OH$ | –7 (300 K) | - | -3 (400 K) | +10 (900 K) | 356 | ref 34 |

[a] $(k_{RPMD} - k_{Quantum\ Mechanical})/k_{Quantum\ Mechanical} \times 100$.
[b] crossover temperature for a given PES used in the calculations.
[c] reciprocal crossover temperature, in a. u. See Ref. 59 for more detail.

**Table 2. Percentage Deviations of the RPMD Thermal Rate Coefficients for Various Prototype Insertion Atom-Diatom Chemical Reactions from the More Accurate Quantum Mechanical Results**

| System | % Error[a] | Reference |
|---|---|---|
| $N(^2D) + H_2 \rightarrow NH + H$ | +(12-9)<br>(270 K ≤ $T$ ≤ 400 K) | ref 40 |
| $O(^1D) + H_2 \rightarrow OH + H$ | −(11-9)<br>(300 K ≤ $T$ ≤ 400 K) | ref 40 |
| $C(^1D) + H_2 \rightarrow CH + H$ | +5<br>($T$ = 300 K) | ref 41 |
| $S(^1D) + H_2 \rightarrow SH + H$ | +9<br>($T$ = 300 K) | ref 41 |

[a] $(k_{\text{RPMD}}\text{-}k_{\text{Quantum Mechanical}})/k_{\text{Quantum Mechanical}} \times 100$.

**Table 3. Comparison of Kinetic Isotope Effects (KIEs) for the $OH + CH_4 \rightarrow CH_3 + H_2O$ Reaction Calculated using Several Theoretical Methods**

| | | TST approaches: | | | | | Experiment |
|---|---|---|---|---|---|---|---|
| *T* (K) | RPMD[a] | cd-QTST[b] | CUS[c] | QI[d] | VTST[e] | CVT/SCT[f] | |
| | | $OH + CH_4 \rightarrow CH_3 + H_2O$ to $OH + CD_4 \rightarrow CD_3 + HDO$ | | | | | |
| 300 | 13.78 | 19.15 | 3.97 | 16.81 | 3.27 | 8.27 | 7.36 (298 K)[g]<br>6.75 (293 K)[h] |
| 400 | 6.38 | 9.00 | 2.45 | 7.54 | 2.43 | 4.82 | 4.31[g]<br>4.04 (409 K)[h] |
| 600 | 3.08 | 3.90 | 1.49 | 4.16 | 1.64 | | 2.63 (602 K)[h] |
| 700 | 2.41 | 3.07 | | 3.12 | 1.52 | | 2.31 (704 K)[h] |
| 800 | 2.16 | 2.64 | 1.45 | 2.80 | | 2.16 | 1.96[h] |
| | | $OH + {}^{12}CH_4 \rightarrow {}^{12}CH_3 + H_2O$ to $OH + {}^{13}CH_4 \rightarrow {}^{13}CH_3 + H_2O$ | | | | | |
| 300 | 0.997 | | | | 1.036 | | 1.005<br>(273-353 K)[i] |

[a] from ref 24 for the $CH_4/CD_4$ KIEs and from ref 45 for the $^{12}C/^{13}C$ KIE.
[b] from ref 24.
[c] from ref 99; μOMT correction is applied.
[d] from ref 100.
[e] from refs 101-103 for the $CH_4/CD_4$ KIEs (ISPE correction is applied) and from ref 45 for the $^{12}C/^{13}C$ KIE (MT correction is applied).
[f] from refs 104,105; SCT correction is applied.
[g] from ref 106.
[h] from ref 107.
[i] from ref 108.

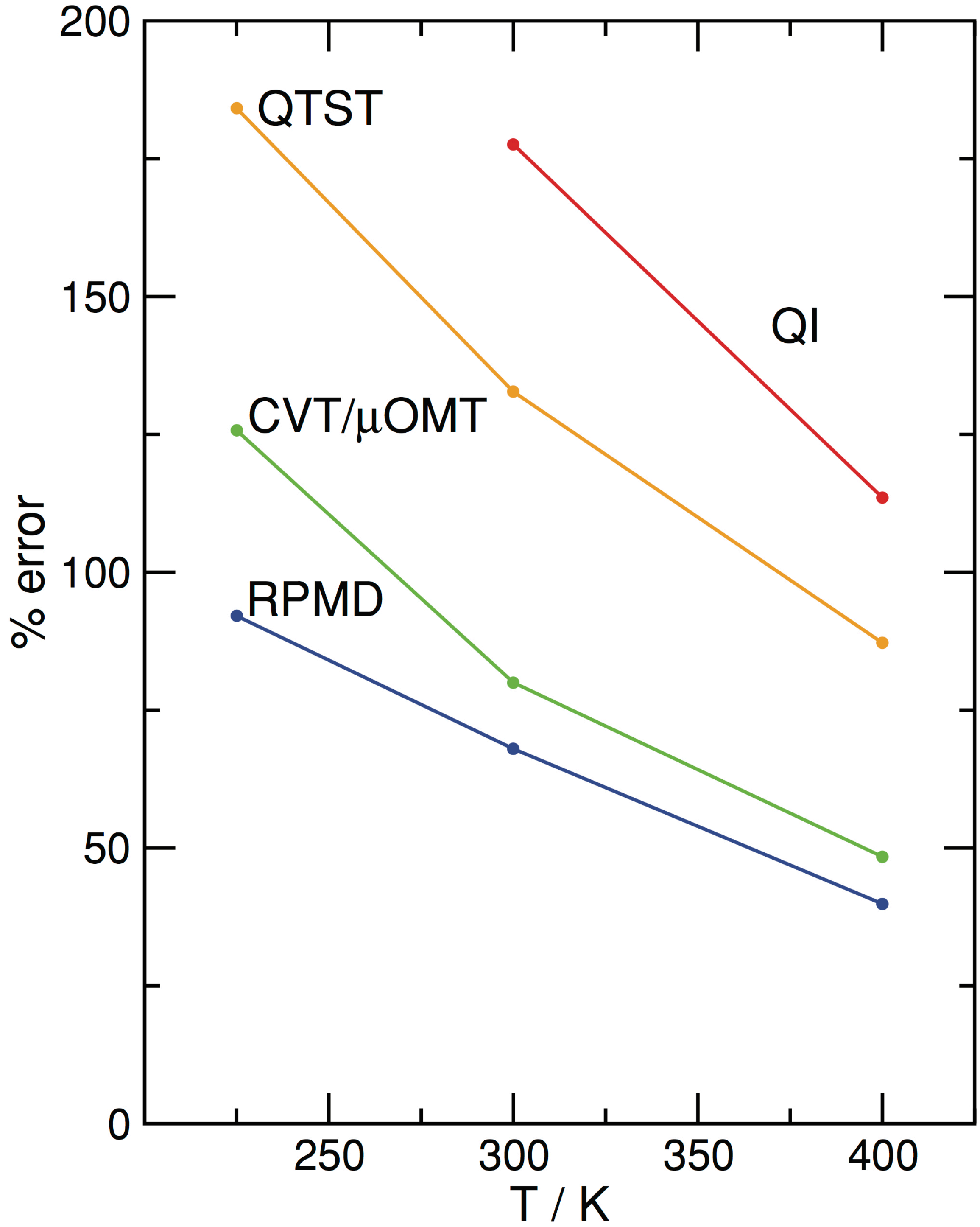

**Figure 1.** Percentage deviations of the quantum instanton (QI), centroid density quantum transition state theory (QTST), canonical variational theory with microcanonically optimized multidimensional tunneling correction (CVT/μOMT), and RPMD rate coefficients for the H + $CH_4$ reaction from the more accurate multiconfiguration time-dependent Hartree (MCTDH) results in the temperature range from 225 to 400 K. Adapted with permission from Suleimanov, Y. V.;

Collepardo-Guevara, R.; Manolopoulos, D. E. Bimolecular reaction rates from ring polymer molecular dynamics: Application to H + $CH_4$ to $H_2$ + $CH_3$. *J. Chem. Phys.* **2011**, *134*, 044131. Copyright 2011, American Institute of Physics.

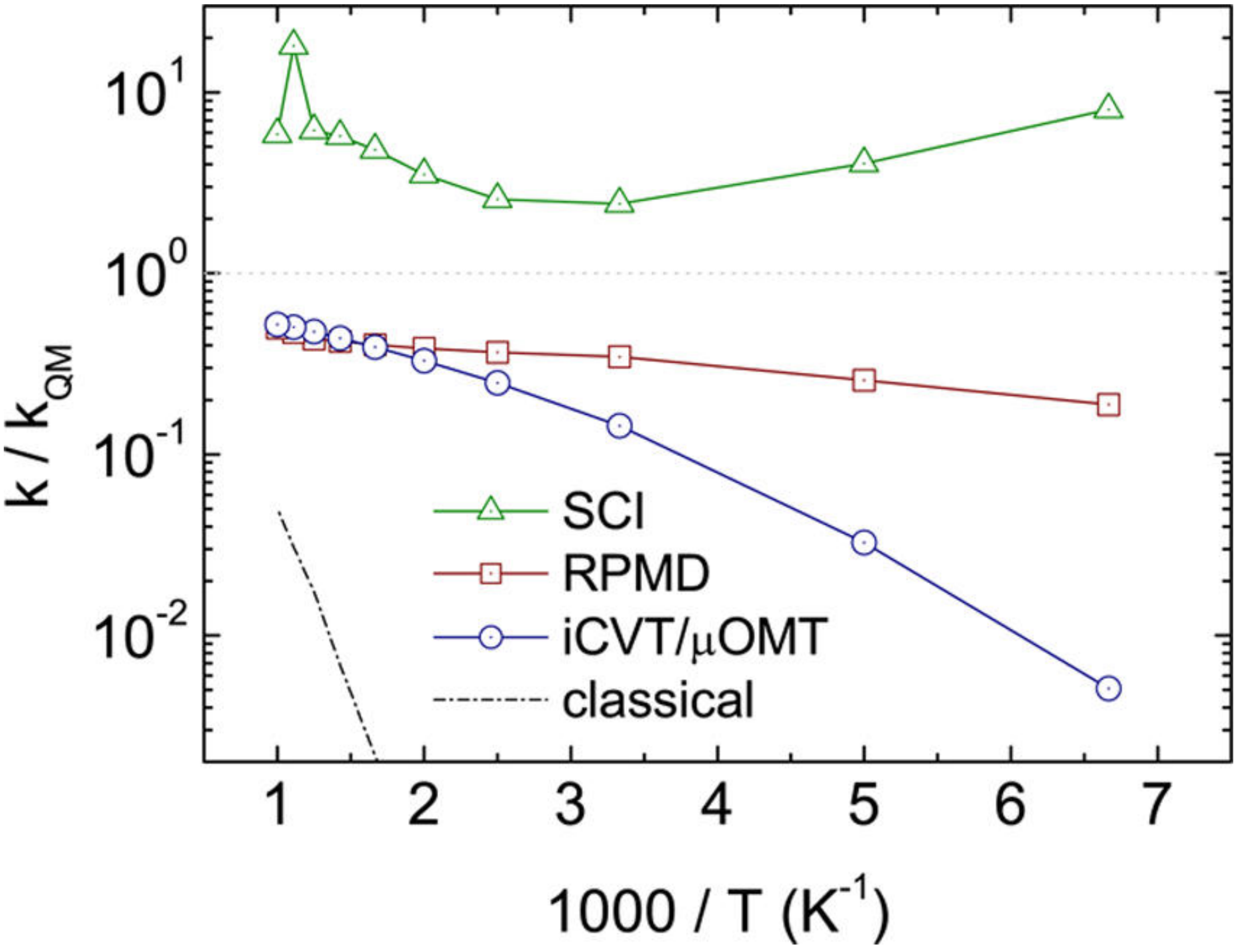

**Figure 2.** Quotient between the computed rates (classical, iCVT/μOMT, where the letter "i" stands for improved treatment of the energetic reaction threshold region, semiclassical instanton (SCI), and RPMD) and the exact QM rates for the D + HMu → DMu + H reaction between 150 and 1000 K. Adapted from Pérez de Tudela, R.; Suleimanov, Y. V.; Richardson, J. O.; Sáez Rábanos, V.; Green, W. H.; Aoiz, F. J. Stress

Test for Quantum Dynamics Approximations: Deep Tunneling in the Muonium Exchange Reaction D + HMu → DMu + H. *J. Phys. Chem. Lett.* **2014**, *5*, 4219–4224. Copyright 2014, American Chemical Society.

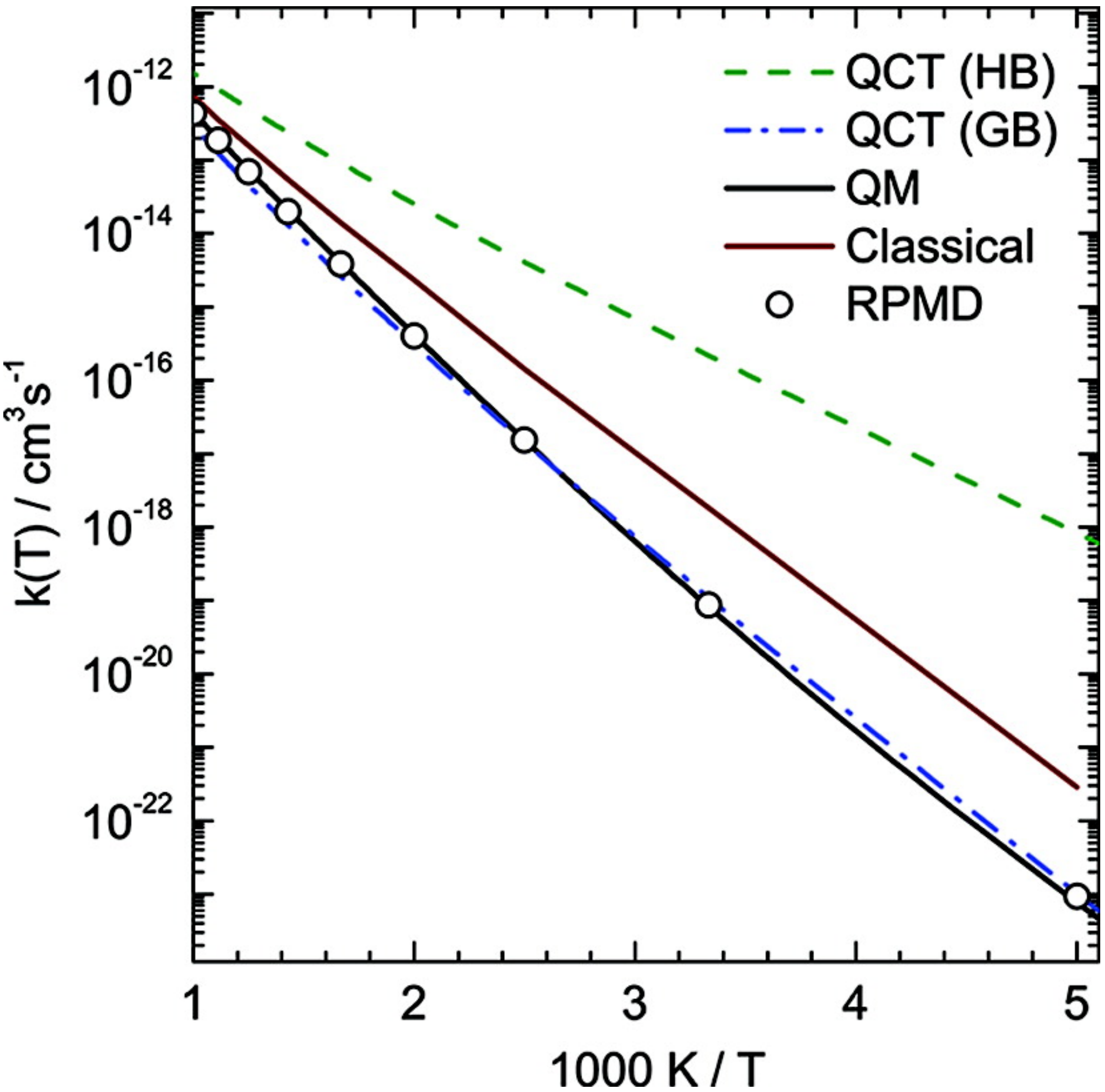

**Figure 3.** Arrhenius plot of quantum mechanical (QM), RPMD, classical, and quasiclassical trajectory (QCT) with histogram (HB) and Gaussian binning (GB) rate coefficients for the Mu+$H_2$ reaction between 200 and 1000 K. Adapted from Perez de Tudela, R.; Aoiz, F. J.; Suleimanov, Y. V.; Manolopoulos, D. E. Chemical reaction rates from ring polymer molecular dynamics: Zero-point energy conservation in Mu + $H_2$ to MuH + H. *J. Phys. Chem. Lett.* **2012**, *3*, 493-497. Copyright 2012, American Chemical Society.

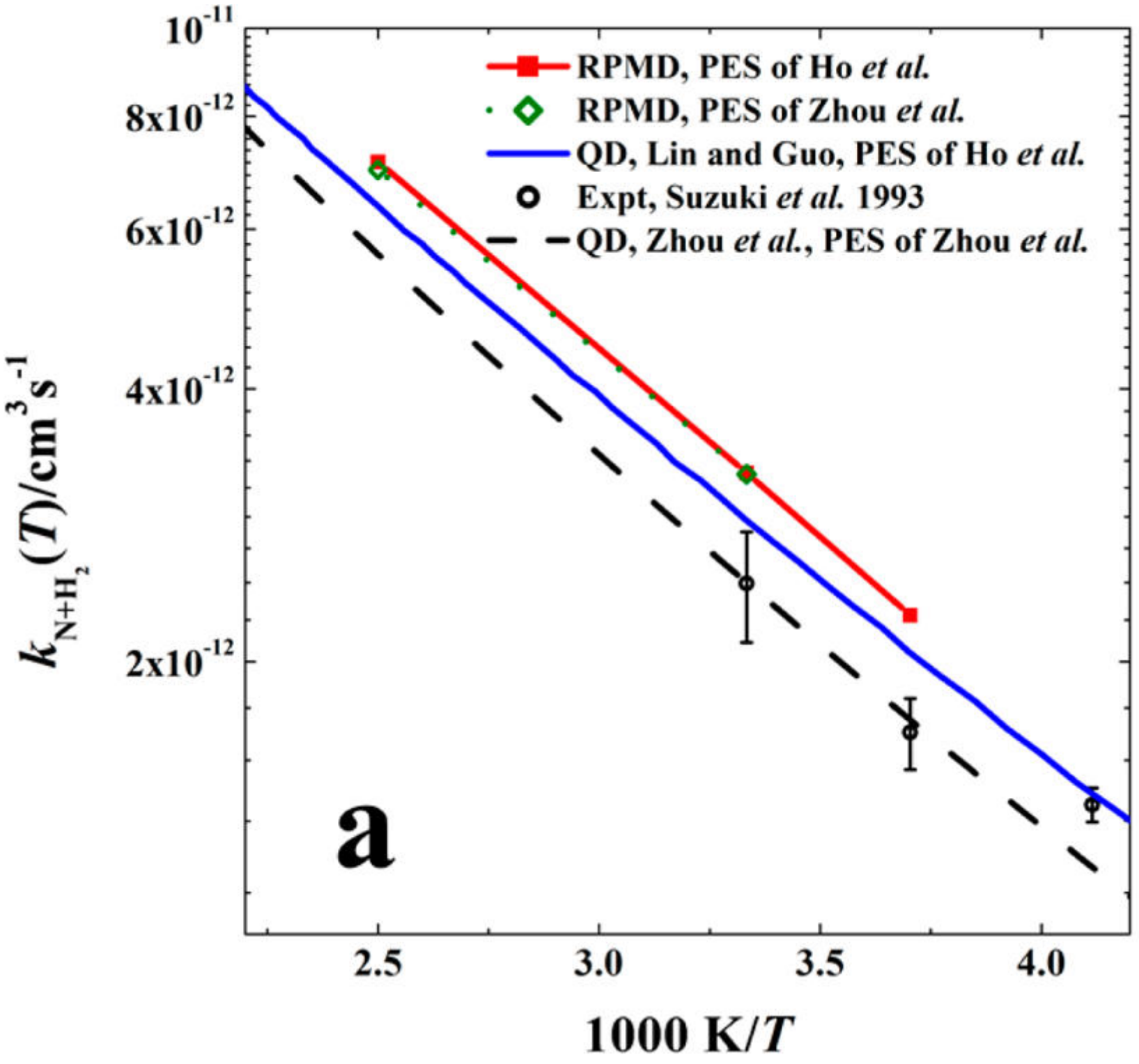

RPMD, PES of Ho et al.
RPMD, PES of Zhou et al.
QD, Lin and Guo, PES of Ho et al.
Expt, Suzuki et al. 1993
QD, Zhou et al., PES of Zhou et al.
a
$k_{N+H_2}(T)$/cm$^3$s$^{-1}$
1000 K/T

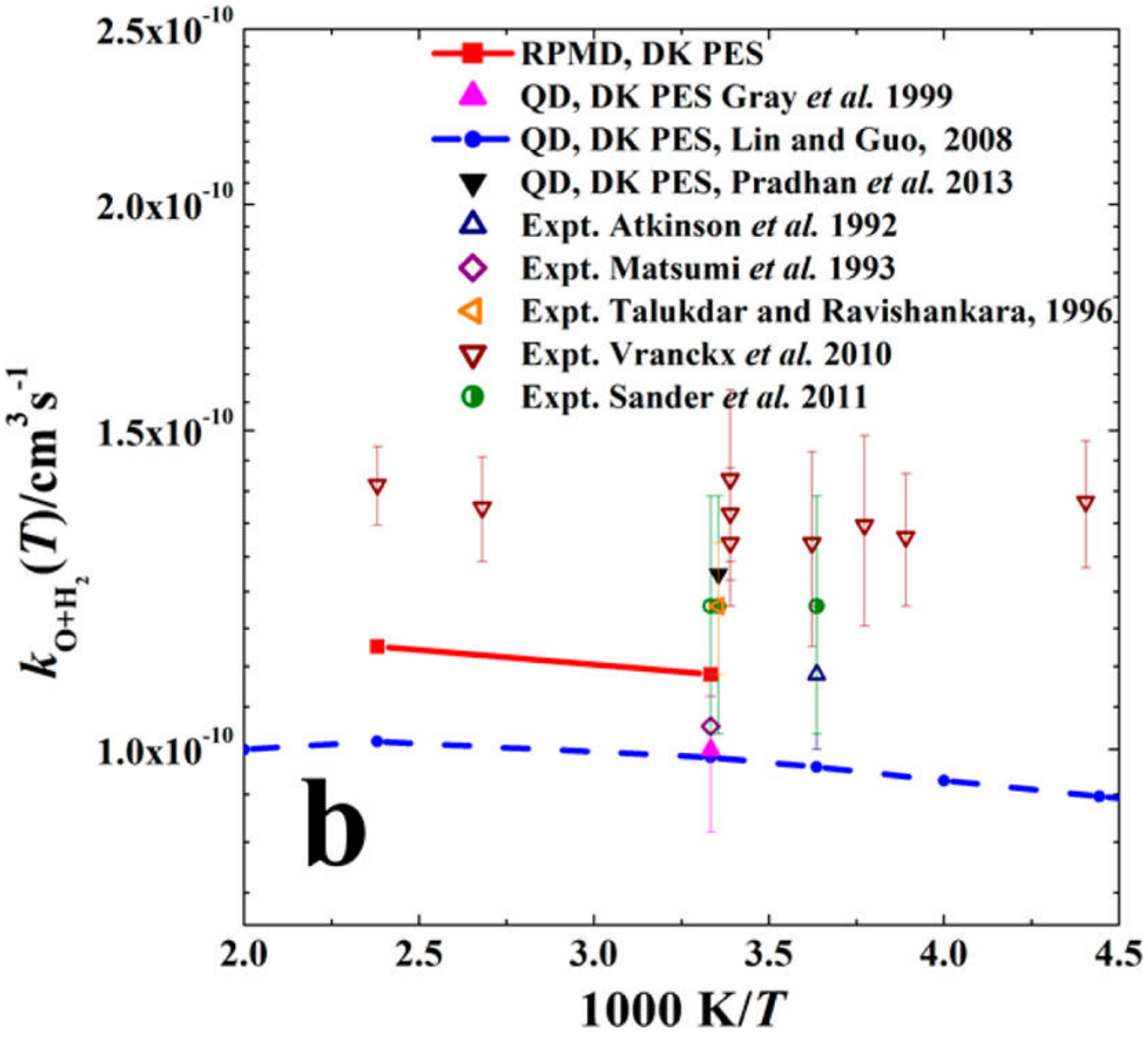

RPMD, DK PES
QD, DK PES Gray et al. 1999
QD, DK PES, Lin and Guo, 2008
QD, DK PES, Pradhan et al. 2013
Expt. Atkinson et al. 1992
Expt. Matsumi et al. 1993
Expt. Talukdar and Ravishankara, 1996
Expt. Vranckx et al. 2010
Expt. Sander et al. 2011
b
$k_{O+H_2}(T)$/cm$^3$s$^{-1}$
1000 K/T

**Figure 4** Comparison of the RPMD, quantum dynamics (QD), and experimental rate coefficients for the N($^2$D) + $H_2$ (upper panel a) and O($^1$D) + $H_2$ (lower panel b) reactions. Adapted from Li, Y.; Suleimanov, Y. V.; Guo, H. Ring-Polymer Molecular Dynamics Rate Coefficient Calculations for Insertion Reactions: X + $H_2$ → HX + H (X = N, O). *J. Phys. Chem. Lett.* **2014**, *5*, 700–705. Copyright 2014, American Chemical Society.

**Biographies**

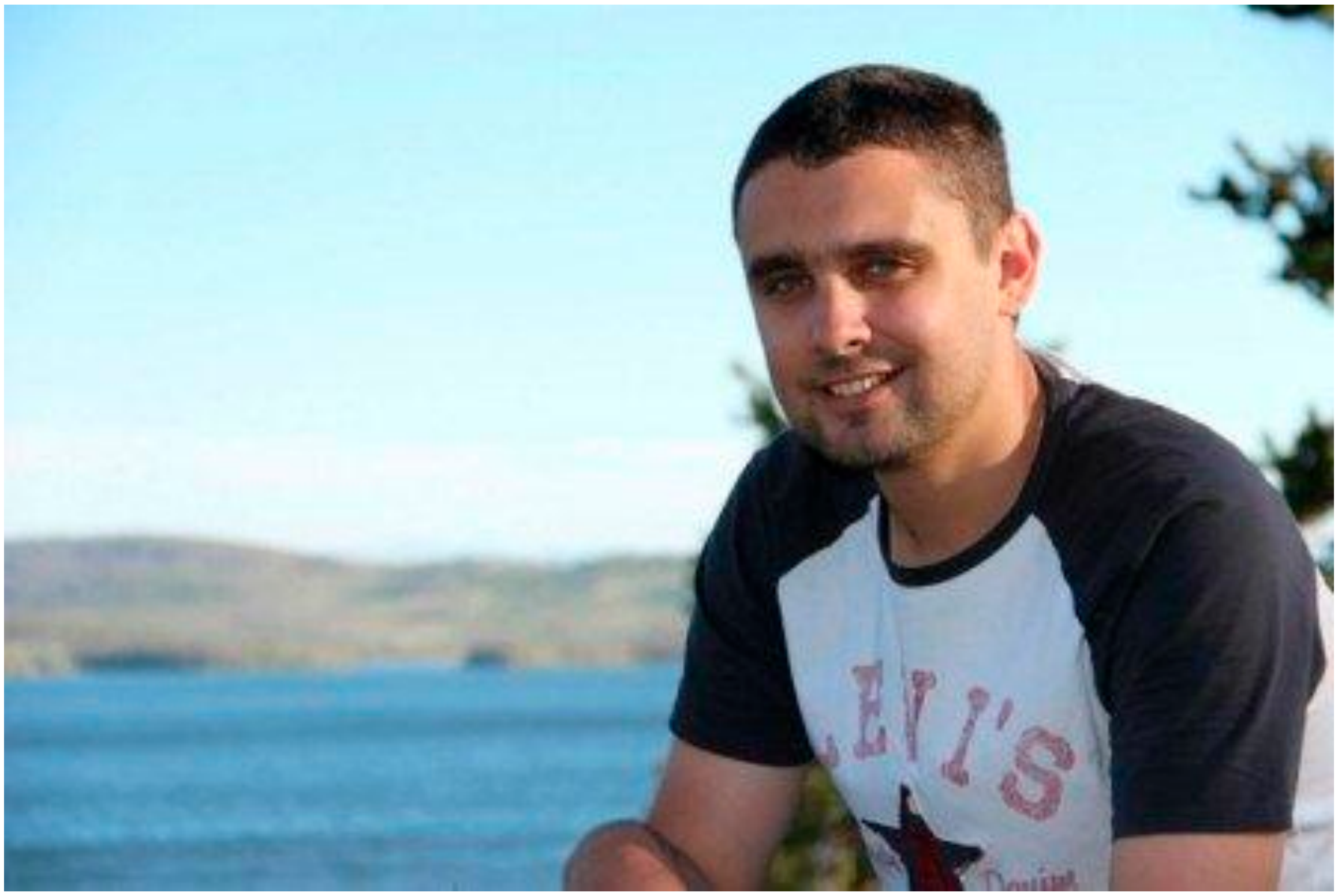

Yury V. Suleimanov received his MSc and PhD degrees from Moscow State University in 2005 and 2009, respectively, under supervision of Prof. Alexei A. Buchachenko. In 2009, he joined the Department of Chemistry, University of Oxford, where he worked in the group of Prof. David E. Manolopoulos as a Royal Society Newton International Fellow. In 2011, Yury became a Princeton Combustion Energy Frontier Research Fellow and worked at the Department of Chemical Engineering, Massachusetts Institute of Technology with Prof. William H. Green. In 2014, Yury was appointed as Assistant Professor at the Cyprus Institute. His research interests are in area of computational chemistry & chemical engineering with focus on the development of the next generation quantum chemistry and dynamics approaches to various chemical kinetics and dynamics problems.

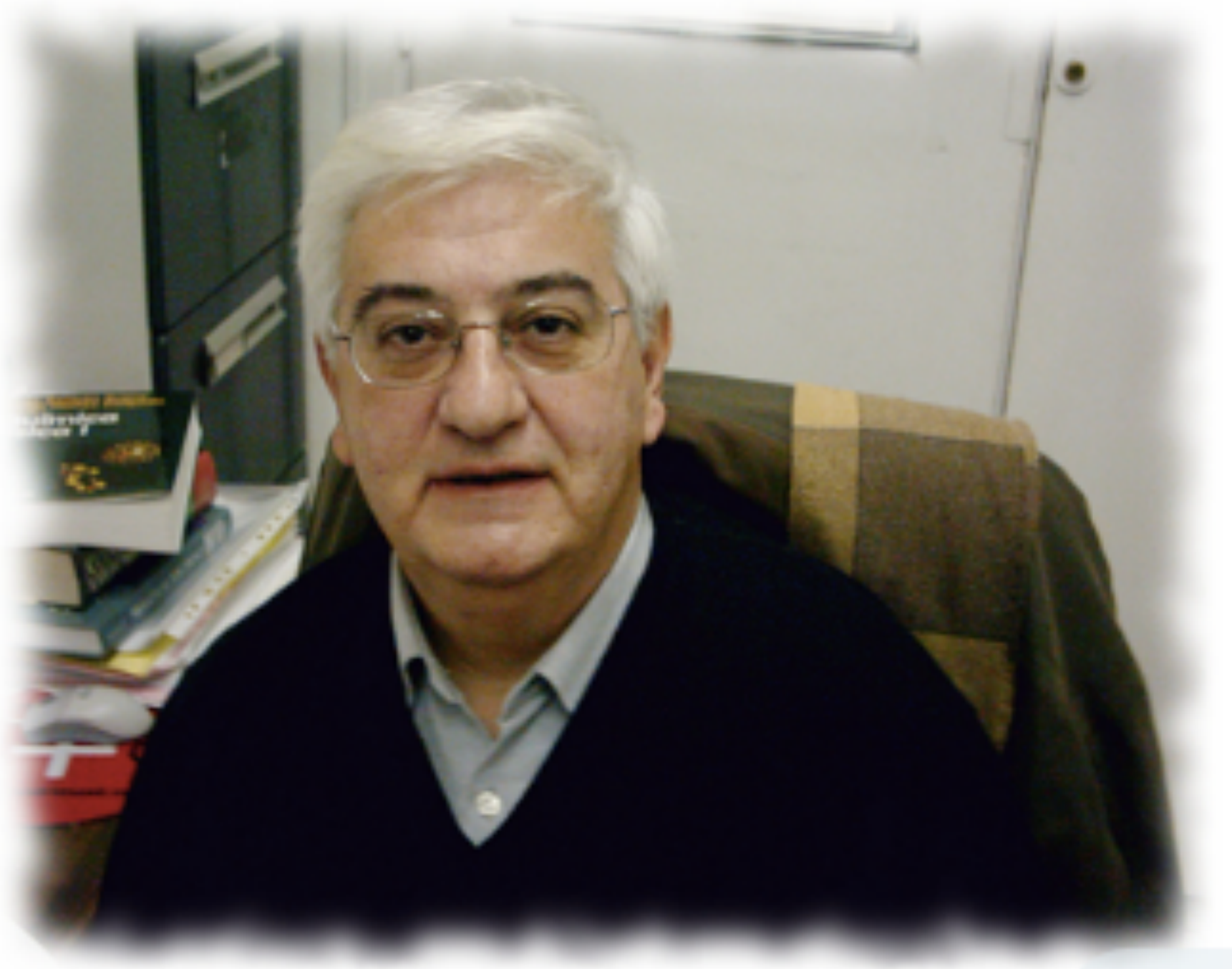

F. Javier Aoiz received his B.S. degree in Chemistry in 1976 from Universidad Complutense de Madrid, Spain and his Ph.D. from that same university in 1981. He spent a two-year period at Columbia University as a Fulbright fellow, working with the late Professor R. B. Bernstein. He joined the faculty of the Physical Chemistry Department at Universidad Complutense de Madrid in 1984 as an Associate Professor and got a chair in Physical Chemistry as a full Professor in 1999. He has been a visiting Professor at the University of Oxford (U.K.) and UCLA (CA, U.S.A.). His research activities are related to experimental and theoretical chemical reaction dynamics and photodissociation processes. His present work is focused on fundamental aspects of reaction dynamics and stereodynamics from both the theoretical and experimental points of view.

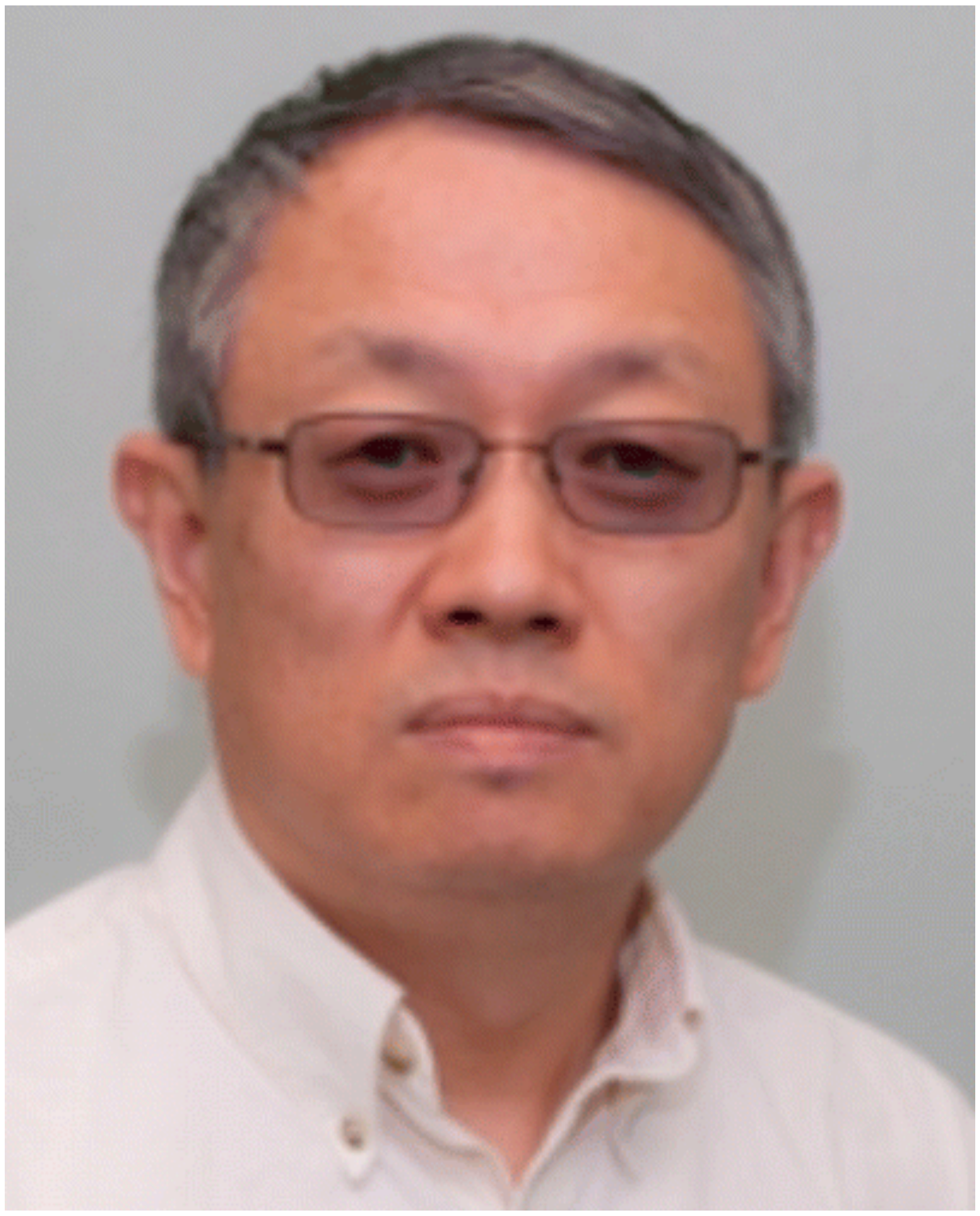

Hua Guo did his undergraduate study at Chengdu Institute of Electronic Engineering, China. After receiving a M.S. degree with Prof. Guo-sen Yan at Sichuan University, China, he moved in 1985 to the U.K. to pursue his D.Phil. degree at Sussex University under the late Prof. John N. Murrell, FRS. Following a postdoctoral appointment with Prof. George C. Schatz at Northwestern University, in 1988–1990, he started his independent career at the University of Toledo in 1990. He moved to the University of New Mexico in 1998 and is now Distinguished Professor of Chemistry and of Physics. He was elected Fellow of the American Physical Society in 2013. He currently serves as a Senior Editor for the Journal of Physical Chemistry A/B/C. His research interests include kinetics and dynamics of reactions in the gas phase and at gas–surface interfaces, heterogeneous catalysis, and enzymatic reactions.

**TOC graphic**

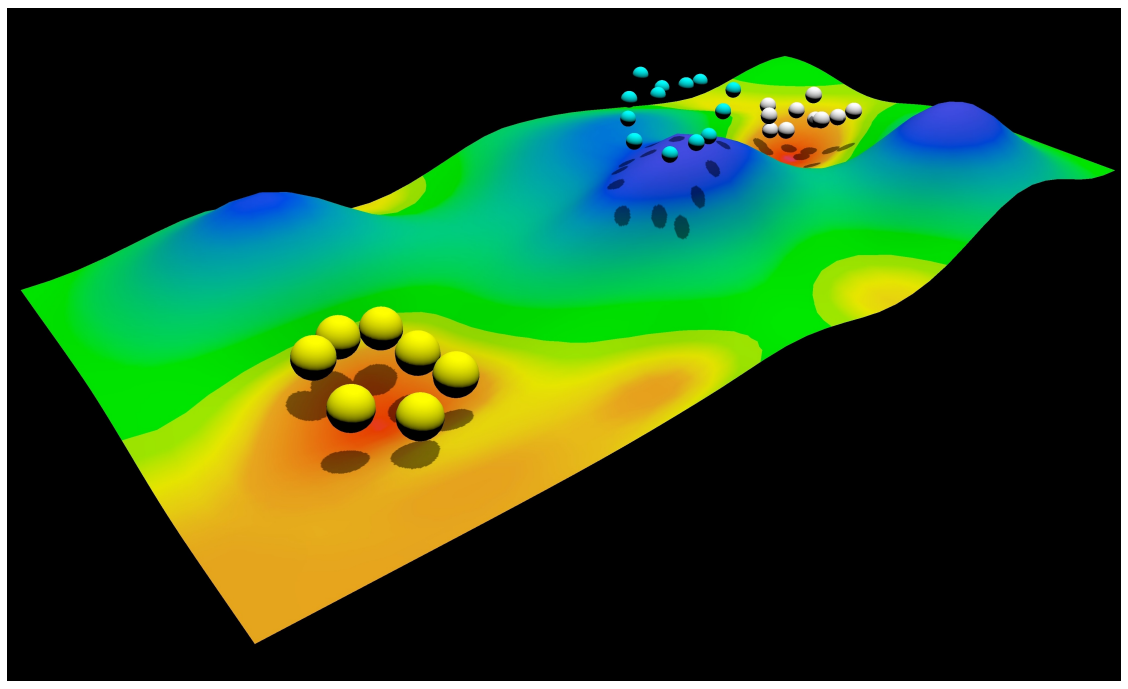